\documentclass[a4paper,11pt]{article}
\usepackage{pos}
\usepackage[absolute,overlay]{textpos}

\title{Imprint of the adjoint meson spectrum in the decay patterns of hidden-bottom tetraquarks}

\author*[a]{Sipaz Sharma}
\author[b]{Juan Andrés Urrea-Niño}
\author[a,d,e]{Nora Brambilla}
\author[c]{Francesco Knechtli}
\author[b]{Michael Peardon}

\affiliation[a]{Physik Department, Technische Universit\"at M\"unchen, James-Franck-Stra{\ss}e 1, D-85748 Garching~b.~M\"unchen, Germany}

\affiliation[b]{School of Mathematics, Trinity College Dublin, Dublin 2, Ireland}
\affiliation[c]{Department of Physics, University of Wuppertal, Gau{\ss}stra{\ss}e 20, 42119 Wuppertal, Germany}
\affiliation[d]{Institute for Advanced Study, Technische Universit\"at M\"unchen, Lichtenbergstra{\ss}e 2 a, D-85748 Garching~b.~M\"unchen, Germany}
\affiliation[e]{Munich Data Science Institute, Technische Universit\"at M\"unchen, Walther-von-Dyck-Stra{\ss}e 10, D-85748 Garching~b.~M\"unchen, Germany}

\emailAdd{sipaz.sharma@tum.de}

\abstract{We aim to clarify the experimentally observed near-degeneracy and decay patterns of the isospin, $I=1$, hidden-bottom tetraquarks $Z_b(10610)$ and $Z_b(10650)$ with quantum numbers $J^{P}=1^{+}$.
	We refer to them as $Z_b$ and $Z_b^{'}$, respectively. In particular, we find first evidence that the suppression of the decay of $Z_b^{'}$ to $B\bar{B^*}$ can be understood in the context of the Born-Oppenheimer Effective Field Theory (BOEFT). BOEFT enables writing both $Z_b$ and $Z_b^{'}$ as superpositions of $Z_1$ and $Z_2$ tetraquark configurations. This decomposition naturally relates the decay patterns of $Z_b$ and $Z_b^{'}$ to the degeneracy of the light degrees of freedom associated with $Z_1$ and $Z_2$ tetraquarks, 
	{\it i.e.,} $1^{--}$and  $0^{-+}$ adjoint mesons, respectively. By calculating the adjoint meson correlators within the framework of lattice QCD, we get good indications that these adjoint mesons are degenerate.}

\FullConference{The 42nd International Symposium on Lattice Field Theory (LATTICE2025)\\
2-8 November 2025\\
Tata Institute of Fundamental Research, Mumbai, India\\}

\begin{document}

\begin{textblock*}{9cm}(14cm,3cm) % adjust position
	\raggedleft
	TUM-EFT 210/26
\end{textblock*}
\maketitle
\section{Introduction}
\subsection{Motivation}
The discovery of $X(3872)$ two decades ago by the Belle experiment in Japan \cite{Belle:2003nnu} ushered in a new era of exotic spectroscopy. $X(3872)$ is an exotic hadron with unconventional quark content: hidden charm ($\bar{c}c$) and two light quarks ($\bar{q} q$) forming an isospin $I=0$ state. Such exotic hadrons with two quarks and two antiquarks are classified as tetraquarks. These hadrons are neither baryons nor mesons, and therefore provide a unique avenue for understanding and exploring properties of Quantum Chromodynamics (QCD), such as confinement and the arrangement of quarks inside color-neutral hadrons. We note that a tetraquark could have one heavy quark-antiquark pair, $[\bar{Q}Q\bar{q}q]$, or two heavy quarks, $[QQ\bar{q}\bar{q}]$. However, for this study, the former configuration is relevant.

Following this, in the year 2011, the Belle experiment discovered two hidden bottom tetraquarks ($\bar{b}b\bar{q}q$): $Z_b(10610)$ and $Z_b^{'}(10650)$ with $I=1$ \cite{Belle:2011aa, Belle:2014vzn}. $Z_b$ lies slightly above $B\bar{B}^*$ threshold and  $Z_b^{'}$ lies slightly above $B^{*}\bar{B}^{*}$. However, the dominant decay channel of $Z_b^{'}$ is $B^{*}\bar{B}^{*}$. In other words, the suppression of $Z_b^{'}$ decay to $B\bar{B}^{*}$ is a puzzle and needs theoretical input. For more details on the current understanding of the exotic hadrons containing $b$ quarks, please refer to  recent reviews \cite{Dai:2026fkg,Brambilla:2019esw}, and references therein.

Since QCD is non-perturbative at the hadronic scale, the experimental progress in the field of states consisting of two heavy quarks ($QQ$) and one pair of heavy quark and antiquark ($\bar{Q}Q$)—a.k.a. XYZ exotics—has, over the years, received important theoretical input from calculations within the framework of lattice QCD.  A distinct approach uses an effective 
field formulation called Born–Oppenheimer Effective Field Theory (BOEFT), which provides a unified framework to explain the spectrum of XYZ exotics and quarkonium \cite{Berwein:2024ztx,Berwein:2015vca}
and needs only a few universal nonperturbative correlators 
as lattice input. We note that the existing lattice QCD results are consistent with the BOEFT framework. In particular, the observed short-distance behaviour of the static energies of hybrids consisting of one heavy quark–antiquark pair bound to gluonic degrees of freedom ($\bar{Q}Qg$) \cite{Juge:2002br,Capitani:2018rox, Hollwieser:2025nvv}, and the observation of string breaking in the $I=0$ quarkonium ($\bar{Q}Q$) channel into two heavy-light mesons ([$\bar{Q}q$][$Q\bar{q}$]) when $Q$ and $\bar{Q}$ are separated by an interquark distance $r\approx 1.2$ fm \cite{Bali:2005fu, Bulava:2024jpj}, can be understood within the unified framework of BOEFT \cite{Berwein:2024ztx}.

\subsection{Application of BOEFT}
BOEFT exploits the energy hierarchy between heavy quarks and the light degrees of freedom (LDF: light quarks and gluons) of QCD. The heavy quarks are static and supply a color source, while the different configurations of the LDF identify different static energies and are characterised by the quantum numbers of the LDF. In addition to isospin and baryon number,  the most important quantum number is the projection of the total angular momentum $\vec{K}$ of the LDF along the axis joining two static color sources, and is labeled as $k^{PC}$, where $k(k+1)$ are the eigenvalues of $|\vec{K}|^2$. We note that for hidden-bottom tetraquarks, LDF comprises of a pair of light quark-antiquark, and thus its baryon number is $0$. The static energies are functions of $r$ and are labeled by BO quantum numbers, $\Lambda_\eta^\sigma$, which are representations of the cylindrical symmetry group. BO quantum numbers are good quantum numbers in the static limit, whereas $k^{PC}$ are good quantum numbers only at short distances where the spherical symmetry gets recovered. $\Lambda_\eta^\sigma$  is directly related to $k^{PC}$: $\Lambda\leq k$ and is represented by Greek letters, $\Sigma, \Pi, ...$, when it takes values $0, 1, ...$; for an exotic state containing $\bar{Q}Q$, $\eta$ is denoted by $g$ and $u$ when $CP$ is $+1$ and $-1$, respectively; and $\sigma$ is the eigenvalue of the reflection operator with respect to a plane passing through the static quark-antiquark axis. The static energies labeled by $\Lambda_\eta^\sigma$ can be plugged into the coupled Schr\"odinger equations in the framework of BOEFT to extract the spectrum of XYZ states.

\begin{table*}[h]
\centering
\begin{tabular}{||c|c|c|c|c||}
\hline
%~ & ~ & ~ &\multicolumn{2}{c|}{LCP$_{[a]}$}&\multicolumn{2}{c|}{LCP$_{[b]}$}  \\
Tetraquark & Adjoint meson spin & BO quantum numbers &$\bar{Q}Q$ spin & $J^{PC}$\\ 
configuration & $k^{PC}$ & $\Lambda_\eta^{\sigma}$ & $S_{\bar{Q}Q}^{PC}$ & \\
\hline \hline
$Z_1$ & $1^{--}$ & $\Sigma_{g}^{+}$ & $0^{-+}$ & $1^{+-}$\\
\hline
$Z_2$ & $0^{-+}$ & $\Sigma_{u}^{-}$ & $1^{--}$ & $1^{+-}$\\
\hline
\end{tabular} 
\caption{This table provides LDF or adjoint meson quantum numbers, BO quantum numbers, and static quark spins corresponding to $Z_1$ and $Z_2$ static energies. The light and static quantum numbers of both $Z_1$ and $Z_2$ combine to give quantum numbers of electrically neutral $Z_b$ and $Z_b'$ {\it i.e.} $J^{PC}=1^{+-}$, and this is why $Z_b$ and $Z_b'$ tetraquarks are  superpositions of the $Z_1$ and $Z_2$ tetraquarks if $Z_1$ and $Z_2$ have similar masses.}
\label{tab:qnos}
\end{table*}
For a hidden-bottom tetraquark configuration in the static limit, if the light quark-antiquark pair does not carry orbital angular momentum, $k$ can take two values: $1/2\otimes 1/2=0\oplus1$ such that both LDF configurations carry $P=-1$, whereas they carry $C=+1$ and $-1$, respectively. Note that the former, $0^{+-}$ has only one static energy associated with it {\it i.e.,} $\Sigma_u^{-}$, whereas the later, $1^{--}$ is associated with a  multiplet of static energies {\it i.e.,} $\Sigma_g^{+}, \Pi_g$. Tab.~\ref{tab:qnos} explicitly shows that $\Sigma_u^{-}$ can be combined with a spin $1$ static $\bar{Q}Q$ configuration to form $Z_2$ tetraquark that carries $J^{PC}$ quantum numbers of electrically neutral $Z_b$ and $Z_b'$. However, for LDF carrying $k^{PC}=1^{--}$, only the static energy carrying BO quantum numbers $\Sigma_g^{+}$ can be combined with spin $0$ static $\bar{Q}Q$ configuration to form $Z_1$ tetraquark, which also carries $J^{PC}$ quantum numbers of electrically neutral $Z_b$ and $Z_b'$.   
If the masses of the physical $Z_1$ and $Z_2$ tetraquarks
obtained as a result of solving the coupled Schr\"odinger equations are
close, then it is expected that the LDF configurations associated with $Z_1$ and
$Z_2$ are degenerate. This in turn implies  that the experimentally observed $Z_b$ and $Z_b'$ can be written as
superpositions of $Z_1$ and $Z_2$ that have the same quantum numbers and similar masses. Let us start with the assumption that $Z_1$ and $Z_2$ have similar masses, and this implies their respective LDF configurations are also degenerate. In particular, the electrically neutral  partners of $Z_b$ and $Z_b'$ tetraquarks with $J^{PC}=1^{+-}$ can be written as superpositions of $Z_1$ and $Z_2$ with $1^{--}$ and $0^{-+}$ as LDF quantum numbers, respectively:

\begin{eqnarray}
|Z_b^{'}\rangle &=& \frac{1}{\sqrt{2}}\bigg(|Z_1\rangle-|Z_2\rangle \bigg) \; ,\nonumber
\\
|Z_b\rangle &=& \frac{1}{\sqrt{2}}\bigg(|Z_1\rangle+|Z_2\rangle \bigg) \; ,
\label{eq:zb}
\end{eqnarray}
where, 
\begin{eqnarray}
 |Z_1\rangle &=& |S_{b\bar{b}}=0\rangle|k^{PC}=1^{--}\rangle ,\nonumber
\\
|Z_2\rangle &=& |S_{b\bar{b}}=1\rangle|k^{PC}=0^{-+}\rangle \; .
\label{eq:z1z2}
\end{eqnarray}
%In Eq.~\eqref{eq:z1z2}, the LDF quantum numbers denoted by $K^{PC}$ are relevant for the lattice calculation in the static limit, as different heavy quark spin configurations, $S_{b\bar{b}}$, cannot be distinguished in this limit. 
In the Heavy Quark Spin Symmetry limit, the LDF quantum numbers relevant for $S\text{-wave}+S\text{-wave}$ heavy-light thresholds and compatible with the quantum numbers of $Z_b$ and $Z_b'$ are also $1^{--}$ and $0^{-+}$. There are two such thresholds corresponding to: i) a pseudoscalar meson and a vector meson ($B\bar{B^*}$), and ii) two vector mesons ($B^*\bar{B^*}$). Therefore, these two heavy-light meson pairs can also be written as superpositions of $Z_1$ and $Z_2$ given in Eq.~\eqref{eq:z1z2} as follows,
\begin{eqnarray}
|B\bar{B^*}\rangle &=& \frac{1}{\sqrt{2}}\bigg(|Z_1\rangle+|Z_2\rangle \bigg) \; ,\nonumber
\\
|B^*\bar{B^*}\rangle &=& \frac{1}{\sqrt{2}}\bigg(-|Z_1\rangle+|Z_2\rangle \bigg) \; .
\label{eq:bbarb}
\end{eqnarray}

The suppression of $Z_b'$ decay to $B\bar{B^*}$ implies that $\langle Z_b^{'}|B\bar{B^*}\rangle$ should vanish. By using equations \eqref{eq:zb}, \eqref{eq:z1z2} and \eqref{eq:bbarb},
\begin{eqnarray}
    \langle Z_b^{'}|B\bar{B^*}\rangle&\propto \langle S_{b\bar{b}}=0|S_{b\bar{b}}=0\rangle\langle k^{PC}=1^{--} |k^{PC}=1^{--}\rangle \nonumber\\
    &-\langle  S_{b\bar{b}}=1|S_{b\bar{b}}=1\rangle\langle k^{PC}=0^{-+}|k^{PC}=0^{-+}\rangle  \text{.}
    \label{eq:supp1}
\end{eqnarray}
After imposing Heavy Quark Spin Symmetry, $\langle S_{b\bar{b}}=0|S_{b\bar{b}}=0\rangle=\langle S_{b\bar{b}}=1|S_{b\bar{b}}=1\rangle$, one obtains,
\begin{equation}
\langle Z_b^{'}|B\bar{B^*}\rangle \propto   \langle k^{PC}=1^{--} |k^{PC}=1^{--}\rangle-\langle k^{PC}=0^{-+}|k^{PC}=0^{-+}\rangle \text{.}
\label{eq:supp2}
\end{equation}
The R.H.S. of Eq.~\ref{eq:supp2} is a difference of two norms and thus vanishes. This means that $\langle Z_b' \mid B \bar{B}^* \rangle$ vanishes only when $Z_1$ and $Z_2$ have similar masses, which in turn implies the degeneracy of LDF configurations (adjoint mesons) associated with the $Z_1$ and $Z_2$ tetraquarks.
As explained in the next section, the LDF configurations associated with these tetraquarks are called {\it{adjoint mesons}}. In Ref.~\cite{Voloshin:2016cgm}, M. B. Voloshin attributed this decay pattern to `Light Quark Spin Symmetry' without providing any proof, and BOEFT now places this on a firm theoretical footing. In the context of BOEFT, this implies the degeneracy of vector, $1^{--}$, and pseudoscalar, $0^{-+}$, adjoint mesons \cite{Berwein:2024ztx, Braaten:2024tbm}.\\

Our aim is to clarify, on one hand, the nature of the adjoint mesons or LDF configurations associated with $Z_1$ and $Z_2$ using lattice QCD calculations, and, on the other hand, to validate the compelling reasoning provided by BOEFT in explaining the experimental observation of the near-degenerate $Z_b'$ and $Z_b$ tetraquarks, as well as the suppression of $Z_b'$ decay to $B\bar{B}^{*}$.

\section{Form of the adjoint meson correlator}
In a tetraquark channel, when the interquark distance, $r$, tends to infinity, the two heavy fundamental, $3$ and anti-fundamental $\bar{3}$ sources should form pairs of heavy-light mesons. On the contrary, when $r\rightarrow 0$, $3\otimes\bar{3}$ reduces to a linear combination of color singlet and octet, $1\oplus8$. For $Q\bar{Q}$, with BO quantum numbers denoted by $\Sigma_{g}^{+}$, the BO potential should smoothly connect the attractive color singlet potential at small $r$ to the $(3+\bar{3})-$potential at large $r$. Whereas, the rest $(3+\bar{3})-$potentials at large $r$ should connect to  the repulsive color octet potential at small $r$. To calculate the static tetraquark potentials on the lattice, Ref.~\cite{Berwein:2024ztx} proposed an appropriate interpolator that reproduces the correct short and long distance behaviour, and is given by, 
\begin{equation}
O_{BO}=\bar{Q}(0)\Gamma_1 \; U(0,r/2) \; T^{a} O_{AM}^{a}(r/2) \; U(r/2,r) Q(r) \; .
\end{equation}
In the equation above, the time coordinate, $T$, is fixed and not explicitly written; $U(x,y)$ represents a spatial Wilson line connecting coordinate $x$ to $y$; $\Gamma_1$ is a general Dirac matrix; $T^{a}$ is a matrix in the adjoint representation of $SU(3)$, which means $a=1, 2,.., 8$; $O_{AM}^{a}$ is the LDF configuration, which in this case is an adjoint meson,
\begin{equation}
    O_{AM}^{a}(r/2)=\bar{q}(r/2)\Gamma T^{a} q(r/2)\; .
\end{equation}
Note that as $r\rightarrow 0$, the tetraquark potential, $V_{O_{BO}}$ reduces to $V_{8}+\Lambda_{AM}$, where $V_8=\frac{\alpha_s}{6r}$, and $\Lambda_{AM}$ is the adjoint meson mass associated with the tetraquark channel.
In Eq.~(5.19) of Ref~\cite{Berwein:2024ztx}, a general LDF correlator is given that can be translated to an adjoint meson correlator, and is given in the following equation with explicit source ($T=0$) and sink ($T=t$) time indices:
 \begin{equation}
 C_{ii}(t)=<O_{AM}^{a}(t) \phi_{adj}^{ab}(t,0)  O_{AM}^{b\dagger}(0)>
 \end{equation}
 The adjoint Wilson line makes the correlator gauge invariant and takes the following form that contains a trace over the color indices:
\begin{equation}
    \phi_{adj}^{ab}(t,0) =\text{Tr} [(U(0,t) T^a U(t,0) T^b)]
    \label{eq:adj}
    \end{equation}
The adjoint meson correlator can be written with general $\Gamma$ matrices, $\Gamma_{snk}$ and $\Gamma_{src}$ such that $\Gamma_{src}=\gamma_0\Gamma_{snk}^{\dagger}\gamma_0$. In the following, the spatial index, $R\equiv r/2$, is arbitrary; Greek letters denote Dirac indices, and Latin letters denote color indices. With all indices explicitly written, the correlator, $C_{ii}(t)$, takes the following form:

 \begin{eqnarray}
    <0|&&\Bar{q}^{\alpha_1}_{c_1}(t,R)\; \Gamma^{\alpha_1\beta_1}_{snk} \; T^{a}_{c_1c_2} q^{\beta_1}_{c_2}(t,R)\nonumber \\ &&\phi_{adj}^{ab}(t,0)\nonumber \\ &&\Bar{q}^{\alpha_2}_{c_3}(0,R)\; \Gamma^{\alpha_2\beta_2}_{src} \; T^{b}_{c_3c_4} q^{\beta_2}_{c_4}(0,R)\;\;\;\;\;\;\;|0 > 
    \label{eq:corr}
\end{eqnarray}
As noted in the previous section, the adjoint mesons corresponding to  $Z_1$ and $Z_2$ are vector, $1^{--}$, with $\Gamma_{snk}=\gamma_i$, and pseudoscalar, $0^{-+}$, with $\Gamma_{snk}=\gamma_5$, respectively. 

\section{Lattice setup}

We used an $N_f = 3 + 1$ ensemble, comprised of  $32^3\times 96$ gauge configurations, produced with the openQCD package. The pion mass $m_\pi$ is tuned to $\approx 406$ MeV for these ensembles. This ensemble has open boundary conditions in time. The lattice spacing, $a$, is set to $0.052$ fm. For further details, see Ref.~\cite{Hollwieser:2020qri, tcbk-x4f1}. For the light quark, we used 100 distillation vectors and standard distillation \cite{HadronSpectrum:2009krc} on 4000 gauge configurations.  The eigenvectors were computed from gauge links that have 20 steps of 3D APE smearing \cite{Albanese-1987} with $\alpha = 0.5$. The temporal links are HYP2 smeared \cite{Hasenfratz:2001hp} with parameters $\alpha_1 = 1$, $\alpha_2 = 1$, $\alpha_3 = 0.5$. %For a given time separation, we average over 48 time sources.

The light quark propagators can be expressed within the distillation framework in terms of perambulators, $\tau$, and Laplacian eigenvectors, $v$, as,
\begin{equation}    
\left \langle q^{\beta_1}_{c_2}(t,R)\; \Bar{q}^{\alpha_2}_{c_3}(0,R)\right \rangle =D^{-1}(R;t,0)^{\beta_1\alpha_2}_{c_2,c_3} 
\displaystyle\rightarrow\displaystyle \sum _{k_1,k_2} v^{k_1}_{R,c_2}(t) \tau^{k_1,k_2}_{\beta_1\alpha_2}(t,0) v^{\dagger k_2}_{R,c_3}(0) \text{.}
\label{eq:prop}
\end{equation}
We used the following color Fierz transformation to go from the adjoint representation to the fundamental representation,
\begin{equation}
    T^{a}_{\alpha \beta} T^{a}_{\gamma \delta} = \delta_{\alpha \delta} \delta_{\gamma \beta} - \delta_{\alpha \beta} \delta_{\gamma \delta}/3 \text{ .}
    \label{fierz}
\end{equation}
After performing Wick contractions in \eqref{eq:corr}, we then substitute \eqref{eq:adj}, \eqref{eq:prop} and \eqref{fierz} to finally obtain the adjoint meson correlator to be calculated on the lattice,
\begin{figure}[h]
\centering
\includegraphics[width=0.9\textwidth, height=0.4\textheight]{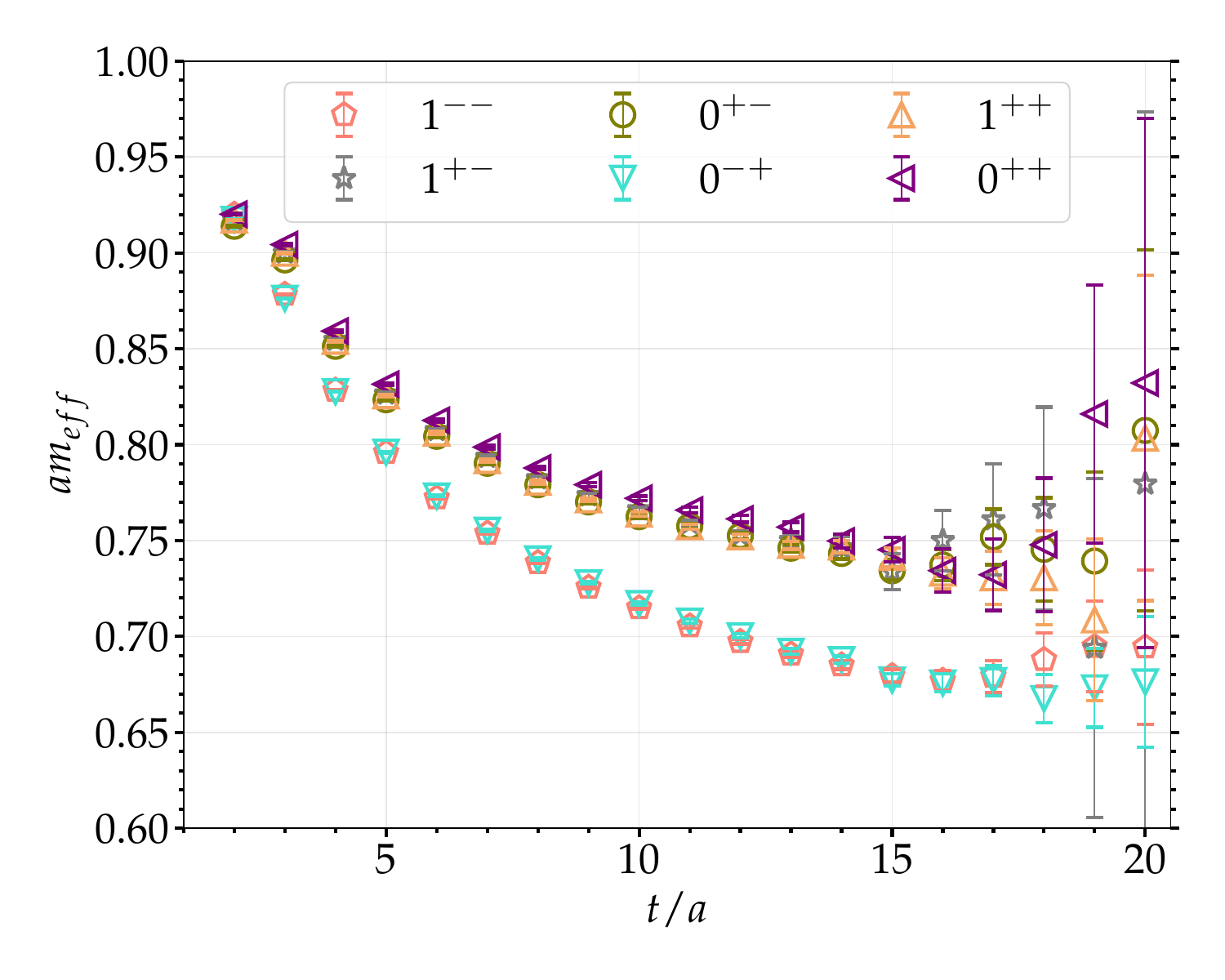}
\caption{This figure shows preliminary effective masses of six $I=1$ adjoint mesons (see text for details). Adjoint mesons with continuum quantum numbers $0^{-+}$ and $1^{--}$ are associated with 
$S\text{-wave}+S\text{-wave}$ heavy-light thresholds, whereas the adjoint mesons carrying continuum quantum numbers $1^{++}$, $0^{+-}$, $0^{++}$ and $1^{+-}$ are associated with $S\text{-wave}+P\text{-wave}$ heavy-light thresholds. The mapping of continuum quantum numbers to their lattice counterparts is given in Tab.~\ref{tab:map}.}
\label{fig:adj}
\end{figure}
\begin{eqnarray}
C_{ii}(t)=\displaystyle\sum_{k_1,k_2,k_3,k_4}&\bigg(  &\Gamma_{src}^{\alpha_2 \beta_2} \tau[0,t]^{k_3 k_4}_{\beta_2 \alpha_1} \hat{\tau}[R;t,0]_{k_4,k_3} 
\Gamma_{snk}^{\alpha_1 \beta_1} \tau[t,0]^{k_1 k_2}_{\beta_1 \alpha_2} \hat{\tau}[R;0,t]_{k_2,k_1} \nonumber \\
    &-&\dfrac{1}{3} \Gamma_{src}^{\alpha_2 \beta_2} \tau[0,t]^{k_3 k_4}_{\beta_2 \alpha_1} \hat{\tau}[R;t,t]_{k_4,k_1} 
\Gamma_{snk}^{\alpha_1 \beta_1} \tau[t,0]^{k_1 k_2}_{\beta_1 \alpha_2} \hat{\tau}[R;0,0]_{k_2,k_3}\bigg)
    \label{eq:final}
    \end{eqnarray}
    In the equation above, we sandwiched the static propagator, which is nothing but a temporal Wilson line, between two Laplacian eigenvectors, and introduced an object, $\hat{\tau}$, called the static perambulator \cite{PhysRevD.107.034511}: 
    \begin{equation}
        \hat{\tau}[R;t,0]_{k_4,k_3} = v^{\dagger k_4}_{R,c_1}(t) U(t,0)_{c_1 c_4}v^{k_3}_{R,c_4}(0)
    \end{equation}
    This implies that when both time indices of the static perambulator are the same, the corresponding temporal Wilson line is identity, and the expression for $\hat{\tau}$ becomes simpler. More specifically, in Eq.~\ref{eq:final},
    \begin{eqnarray}
     \hat{\tau}[R;t,t]_{k_4,k_1}&=v^{\dagger k_4}_{R,c_1}(t) v^{k_1}_{R,c_1}(t) \text{ ,}  \nonumber\\
     \hat{\tau}[R;0,0]_{k_2,k_3}&=v^{\dagger k_2}_{R,c_1}(0) v^{k_3}_{R,c_1}(0) \text{ .}
    \end{eqnarray}
    
\section{Results and conclusions}
We extracted effective masses, $am_{eff}$, from various adjoint meson correlators calculated using Eq.~\eqref{eq:final}. All the statistical error analysis is done using the pyerrors library \cite{Joswig2023} which uses the $\Gamma$-method \cite{Wolff2004, Wolff2007, Schaefer2011} with automatic differentiation \cite{Ramos2019} to systematically account for the correlation between samples in the Monte-Carlo data. Note that in the following, we explicitly write the lattice units. As is well known,
\begin{equation}
am_{eff}(t/a,t/a+1)=\text{log }\dfrac{C_{ii}(t/a)}{C_{ii}(t/a+1)} \text{ ,}
\end{equation}
and because of the spectral decomposition,
%\begin{equation}
%    C_{ij}(t/a)=\sum_{n=1}^{\infty} e^{-E_{n}t}\langle\hat{O}_{i}|n\rangle\langle n|\hat{O}_j\rangle \text{,}
%\end{equation}
in the large $t$ limit, it is expected that the contamination from the excited states will reduce, and  $am_{eff}$ will plateau to give the ground state in each quantum number channel. However, as can be seen from Fig.~\ref{fig:adj}, in all adjoint meson channels, the signal-to-noise ratio deteriorates before a proper plateau can be observed. Nevertheless, even at the level of effective masses, the pseudoscalar, $0^{-+}$, and vector $1^{--}$ adjoint mesons computations are in agreement. To be more specific, for $1^{--}$, so far we only used $\gamma_1$ and not an average over $\gamma_1$, $\gamma_2$ and $\gamma_3$, whereas for $0^{-+}$ we used $\gamma_0\gamma_5$. We emphasise that Fig.~\ref{fig:adj} shows preliminary results. In particular, in the vector channel, a full analysis averaging over the irreducible representations of the cubic group is in progress. The upper block of Tab.~\ref{tab:map} maps the continuum quantum numbers of these two adjoint mesons to their lattice counterparts. The adjoint meson correlators are averaged over all spatial points in the $x$- and $y$-directions while keeping the $z$-coordinate fixed to $0$. For the temporal averaging, in order to avoid the open boundary effects, we restrict the source and sink times to $t_1/a, t_2/a \in [24,71]$, respectively, with the condition $t_2 > t_1$. For a fixed source--sink separation $t/a = t_2/a - t_1/a$, this implies that the number of available time sources is $48 - t/a$, over which the correlators are averaged.
\begin{table*}
\centering
\begin{tabular}{||c|c|c||}
\hline
%~ & ~ & ~ &\multicolumn{2}{c|}{LCP$_{[a]}$}&\multicolumn{2}{c|}{LCP$_{[b]}$}  \\
Continuum quantum numbers & Lattice quantum numbers& $\Gamma_{snk}$ used in the computation \\
\hline \hline
 $0^{-+}$& $A_{1}^{-+}$ & $\gamma_0\gamma_5$  \\
 \hline
 $1^{--}$& $T_{1}^{--}$ & $\gamma_1$  \\ 
\hline \hline
$1^{++}$& $T_{1}^{++}$ & $\gamma_5\gamma_1$  \\ 
\hline
$0^{+-}$& $A_{1}^{+-}$ & $\gamma_0$  \\ 
\hline
$0^{++}$& $A_{1}^{++}$ & $\mathbb{I}$  \\ 
\hline
$1^{+-}$& $T_{1}^{+-}$ & $\gamma_2\gamma_3$  \\
\hline
$2^{++}$& $E^{++}$, $T_{2}^{++}$ & -  \\
\hline
$2^{+-}$& $E^{+-}$, $T_{2}^{+-}$ & -  \\
\hline
\end{tabular} 
\caption{This table maps the continuum quantum numbers of eight adjoint mesons to their lattice quantum numbers based on the irreducible representations of the cubic group \cite{JOHNSON1982147}. The third column provides explicit $\Gamma_{snk}$ matrices given as input to the correlator in Eq.~\ref{eq:final} to produce $am_{eff}$ for each quantum number channel shown in Fig.~\ref{fig:adj}. }
\label{tab:map}
\end{table*}

Our results on the spectrum of adjoint mesons associated with $Z_1$ and $Z_2$,  indicate that the decay patterns of hidden-bottom tetraquarks  that lie close to $S\text{-wave}+S\text{-wave}$ heavy-light thresholds can be attributed to the degeneracy of the adjoint mesons. We note that in the quenched case, motivated by the supersymmetric arguments, such a calculation has been performed before by Foster and Michael \cite{Foster:1998wu}. However, their results were plagued by large errors. The vector and pseudoscalar adjoint mesons were found to be $-10(103)$ MeV and $34(161)$ MeV heavier than the lightest gluelump. At a fixed value of the inverse gauge coupling, Ref.~\cite{Foster:1998wu} also studied the degeneracy pattern of these adjoint mesons at two different values of light quark masses. While, as expected, the masses of the adjoint mesons were dependent upon the input light quark masses, the degeneracy within errors turned out to be  independent of the  input light quark mass. Therefore, we expect that  while approaching the chiral limit, the observed degeneracy pattern should persist. 

 In addition to this, we also calculated effective masses, $am_{eff}$, of the adjoint mesons related to the $S\text{-wave}+P\text{-wave}$ heavy-light thresholds. There are six such adjoint mesons with continuum quantum numbers $1^{++}$, $0^{+-}$, $0^{++}$, $1^{+-}$, $2^{++}$ and $2^{+-}$. The lower block of Tab.~\ref{tab:map} maps the continuum quantum numbers of these six adjoint mesons to their lattice counterparts.  In Fig.~\ref{fig:adj}, we show results for the first four adjoint mesons, which are calculated by substituting $\Gamma_{snk}$ with $\gamma_5\gamma_1,\gamma_0,\mathbb{I}, \gamma_2\gamma_3$, respectively.  As can be seen, the results for these four adjoint mesons also agree with each other. All these results point towards the idea of Light Quark Spin Symmetry conceived by Voloshin, and put in a rigorous mathematical framework by BOEFT. 
 
 The current lattice data support the BOEFT interpretation and are
consistent with the degeneracy of the relevant adjoint mesons at the level of the effective masses, but a
definitive statement requires improved plateau control and a  Generalised Eigenvalue Problem (GEVP) analysis. As part of the outlook, our short-term goal is to incorporate distillation profiles \cite{Knechtli:2022bji} and implement the GEVP in each quantum-number channel in order to obtain improved plateaux. We also plan to compute the remaining two adjoint mesons associated with $S\text{-wave}+P\text{-wave}$ thresholds {\it i.e.,}  $2^{++}$ and $2^{+-}$.
\section*{Acknowledgements}

The authors thank Abhishek Mohapatra and Antonio Vairo for the useful discussions. The authors express their gratitude to Roman Höllwieser for the generation of the gauge configurations and perambulators. N.B. and S.S. acknowledge support from the German Research Foundation (DFG) cluster of excellence ORIGINS funded by the Deutsche Forschungsgemeinschaft under Germany’s Excellence Strategy-EXC-2094-390783311. N.B. acknowledges the Advanced ERC grant ERC-2023-ADG-Project EFT-XYZ. J.A.U.-N. acknowledges support from a Research Ireland (Science Foundation Ireland) Frontiers for the Future Project award [grant number SFI-21/FFP-P/10186]. This work is also supported by the DFG research unit FOR5269 "Future methods for studying confined gluons in QCD". The authors gratefully acknowledge the Gauss Centre for Supercomputing e.V. (www.gauss-centre.eu) for funding this project by providing computing time on the GCS Supercomputer SuperMUC-NG at Leibniz Supercomputing Centre (www.lrz.de) as well as the Juelich Supercomputing Centre (JSC) for allocating space on the storage system JUST (project ID hwu35).

%\begin{thebibliography}{99}
%\bibitem{...}

%\end{thebibliography}
\bibliographystyle{unsrt}

\bibliography{refs}

@article{Foster:1998wu,
    author = "Foster, M. and Michael, Christopher",
    collaboration = "UKQCD",
    title = "{Hadrons with a heavy color adjoint particle}",
    eprint = "hep-lat/9811010",
    archivePrefix = "arXiv",
    reportNumber = "LTH-444",
    doi = "10.1103/PhysRevD.59.094509",
    journal = "Phys. Rev. D",
    volume = "59",
    pages = "094509",
    year = "1999"
}

@article{Berwein:2015vca,
    author = "Berwein, Matthias and Brambilla, Nora and Tarr{\'u}s Castell{\`a}, Jaume and Vairo, Antonio",
    title = "{Quarkonium Hybrids with Nonrelativistic Effective Field Theories}",
    eprint = "1510.04299",
    archivePrefix = "arXiv",
    primaryClass = "hep-ph",
    reportNumber = "TUM-EFT-45-14",
    doi = "10.1103/PhysRevD.92.114019",
    journal = "Phys. Rev. D",
    volume = "92",
    number = "11",
    pages = "114019",
    year = "2015"
}

@article{Braaten:2024tbm,
    author = "Braaten, Eric and Bruschini, Roberto",
    title = "{Exotic hidden-heavy hadrons and where to find them}",
    eprint = "2409.08002",
    archivePrefix = "arXiv",
    primaryClass = "hep-ph",
    doi = "10.1016/j.physletb.2025.139386",
    journal = "Phys. Lett. B",
    volume = "863",
    pages = "139386",
    year = "2025"
}

@article{Brambilla:2019esw,
    author = "Brambilla, Nora and Eidelman, Simon and Hanhart, Christoph and Nefediev, Alexey and Shen, Cheng-Ping and Thomas, Christopher E. and Vairo, Antonio and Yuan, Chang-Zheng",
    title = "{The $XYZ$ states: experimental and theoretical status and perspectives}",
    eprint = "1907.07583",
    archivePrefix = "arXiv",
    primaryClass = "hep-ex",
    reportNumber = "TUM-EFT 125/19",
    doi = "10.1016/j.physrep.2020.05.001",
    journal = "Phys. Rept.",
    volume = "873",
    pages = "1--154",
    year = "2020"
}

@article{Hollwieser:2020qri,
    author = {H{\"o}llwieser, Roman and Knechtli, Francesco and Korzec, Tomasz},
    collaboration = "ALPHA",
    title = "{Scale setting for $N_f=3+1$ QCD}",
    eprint = "2002.02866",
    archivePrefix = "arXiv",
    primaryClass = "hep-lat",
    reportNumber = "WUB/20-00",
    doi = "10.1140/epjc/s10052-020-7889-7",
    journal = "Eur. Phys. J. C",
    volume = "80",
    number = "4",
    pages = "349",
    year = "2020"
}

@article{Hollwieser:2025nvv,
    author = {H{\"o}llwieser, Roman and Knechtli, Francesco and Korzec, Tomasz and Peardon, Michael J. and Struckmeier, Laura and Urrea-Ni{\~n}o, Juan Andr{\'e}s},
    title = "{Hybrid static potentials and gluelumps on $N_f=3+1$ ensembles}",
    eprint = "2501.15670",
    archivePrefix = "arXiv",
    primaryClass = "hep-lat",
    doi = "10.22323/1.466.0102",
    journal = "PoS",
    volume = "LATTICE2024",
    pages = "102",
    year = "2025"
}

@article{Bulava:2024jpj,
    author = "Bulava, John and Knechtli, Francesco and Koch, Vanessa and Morningstar, Colin and Peardon, Michael",
    title = "{The quark-mass dependence of the potential energy between static colour sources in the QCD vacuum with light and strange quarks}",
    eprint = "2403.00754",
    archivePrefix = "arXiv",
    primaryClass = "hep-lat",
    reportNumber = "WUB/24-00",
    doi = "10.1016/j.physletb.2024.138754",
    journal = "Phys. Lett. B",
    volume = "854",
    pages = "138754",
    year = "2024"
}

@article{Juge:2002br,
    author = "Juge, K. Jimmy and Kuti, Julius and Morningstar, Colin",
    title = "{Fine structure of the QCD string spectrum}",
    eprint = "hep-lat/0207004",
    archivePrefix = "arXiv",
    doi = "10.1103/PhysRevLett.90.161601",
    journal = "Phys. Rev. Lett.",
    volume = "90",
    pages = "161601",
    year = "2003"
}

@article{Belle:2011aa,
    author = "Bondar, A. and others",
    collaboration = "Belle",
    title = "{Observation of two charged bottomonium-like resonances in Y(5S) decays}",
    eprint = "1110.2251",
    archivePrefix = "arXiv",
    primaryClass = "hep-ex",
    doi = "10.1103/PhysRevLett.108.122001",
    journal = "Phys. Rev. Lett.",
    volume = "108",
    pages = "122001",
    year = "2012"
}

@article{Belle:2014vzn,
    author = "Garmash, A. and others",
    collaboration = "Belle",
    title = "{Amplitude analysis of $e^+ e^- \to \Upsilon(nS) \pi^+ \pi^-$ at $\sqrt{s}=10.865$~GeV}",
    eprint = "1403.0992",
    archivePrefix = "arXiv",
    primaryClass = "hep-ex",
    reportNumber = "BELLE-PREPRINT-2014-4, KEK-PREPRINT-2013-63",
    doi = "10.1103/PhysRevD.91.072003",
    journal = "Phys. Rev. D",
    volume = "91",
    number = "7",
    pages = "072003",
    year = "2015"
}

@article{Belle:2003nnu,
    author = "Choi, S. K. and others",
    collaboration = "Belle",
    title = "{Observation of a narrow charmonium-like state in exclusive $B^\pm \to K^\pm \pi^+ \pi^- J/\psi$ decays}",
    eprint = "hep-ex/0309032",
    archivePrefix = "arXiv",
    doi = "10.1103/PhysRevLett.91.262001",
    journal = "Phys. Rev. Lett.",
    volume = "91",
    pages = "262001",
    year = "2003"
}

@article{Berwein:2024ztx,
    author = "Berwein, Matthias and Brambilla, Nora and Mohapatra, Abhishek and Vairo, Antonio",
    title = "{Hybrids, tetraquarks, pentaquarks, doubly heavy baryons, and quarkonia in Born-Oppenheimer effective theory}",
    eprint = "2408.04719",
    archivePrefix = "arXiv",
    primaryClass = "hep-ph",
    reportNumber = "TUM-EFT 185/23",
    doi = "10.1103/PhysRevD.110.094040",
    journal = "Phys. Rev. D",
    volume = "110",
    number = "9",
    pages = "094040",
    year = "2024"
}

@article{Voloshin:2016cgm,
    author = "Voloshin, M. B.",
    title = "{Light Quark Spin Symmetry in $Z_b$ Resonances?}",
    eprint = "1601.02540",
    archivePrefix = "arXiv",
    primaryClass = "hep-ph",
    reportNumber = "FTPI-MINN-16-01, UMN-TH-3512-16",
    doi = "10.1103/PhysRevD.93.074011",
    journal = "Phys. Rev. D",
    volume = "93",
    number = "7",
    pages = "074011",
    year = "2016"
}

@article{Dai:2026fkg,
    author = "Dai, Xinchen and Jia, Sen and Nefediev, Alexey and Nieves, Juan and Shen, Chengping and Zhang, Liming",
    title = "{Exotic hadrons associated with $b$-quark}",
    note = {arXiv:2603.09315 [hep-ph], 2026}
}

@article{HadronSpectrum:2009krc,
    author = "Peardon, Michael and Bulava, John and Foley, Justin and Morningstar, Colin and Dudek, Jozef and Edwards, Robert G. and Joo, Balint and Lin, Huey-Wen and Richards, David G. and Juge, Keisuke Jimmy",
    collaboration = "Hadron Spectrum",
    title = "{A Novel quark-field creation operator construction for hadronic physics in lattice QCD}",
    eprint = "0905.2160",
    archivePrefix = "arXiv",
    primaryClass = "hep-lat",
    reportNumber = "JLAB-THY-09-985",
    doi = "10.1103/PhysRevD.80.054506",
    journal = "Phys. Rev. D",
    volume = "80",
    pages = "054506",
    year = "2009"
}

@article{Hasenfratz:2001hp,
    author = "Hasenfratz, Anna and Knechtli, Francesco",
    title = "{Flavor symmetry and the static potential with hypercubic blocking}",
    eprint = "hep-lat/0103029",
    archivePrefix = "arXiv",
    reportNumber = "COLO-HEP-462",
    doi = "10.1103/PhysRevD.64.034504",
    journal = "Phys. Rev. D",
    volume = "64",
    pages = "034504",
    year = "2001"
}

@article{Knechtli:2022bji,
    author = "Knechtli, Francesco and Korzec, Tomasz and Peardon, Michael and Urrea-Ni{\~n}o, Juan Andr{\'e}s",
    title = "{Optimizing creation operators for charmonium spectroscopy on the lattice}",
    eprint = "2205.11564",
    archivePrefix = "arXiv",
    primaryClass = "hep-lat",
    reportNumber = "WUB/22-02",
    doi = "10.1103/PhysRevD.106.034501",
    journal = "Phys. Rev. D",
    volume = "106",
    number = "3",
    pages = "034501",
    year = "2022"
}

@article{PhysRevD.107.034511,
  title = {Constructing static quark-antiquark creation operators from Laplacian eigenmodes},
  author = {H\"ollwieser, Roman and Knechtli, Francesco and Korzec, Tomasz and Peardon, Michael and Urrea-Ni\~no, Juan Andr\'es},
  journal = {Phys. Rev. D},
  volume = {107},
  issue = {3},
  pages = {034511},
  numpages = {9},
  year = {2023},
  month = {Feb},
  publisher = {American Physical Society},
  doi = {10.1103/PhysRevD.107.034511},
  url = {https://link.aps.org/doi/10.1103/PhysRevD.107.034511}
}

@article{JOHNSON1982147,
title = {Angular momentum on a lattice},
journal = {Physics Letters B},
volume = {114},
number = {2},
pages = {147-151},
year = {1982},
issn = {0370-2693},
doi = {https://doi.org/10.1016/0370-2693(82)90134-4},
url = {https://www.sciencedirect.com/science/article/pii/0370269382901344},
author = {R.C. Johnson}
}

@article{Albanese-1987,
title = {Glueball masses and string tension in lattice QCD},
journal = {Physics Letters B},
volume = {192},
number = {1},
pages = {163-169},
year = {1987},
issn = {0370-2693},
doi = {https://doi.org/10.1016/0370-2693(87)91160-9},
url = {https://www.sciencedirect.com/science/article/pii/0370269387911609},
author = {M. Albanese and others}
}

@article{tcbk-x4f1,
  title = {S-wave flavor-singlet meson mixing in QCD with light and charm quarks},
  author = {Urrea-Ni\~no, Juan Andr\'es and H\"ollwieser, Roman and Knechtli, Francesco and Korzec, Tomasz and Finkenrath, Jacob and Peardon, Michael},
  journal = {Phys. Rev. D},
  volume = {112},
  issue = {7},
  pages = {074502},
  numpages = {14},
  year = {2025},
  month = {Oct},
  publisher = {American Physical Society},
  doi = {10.1103/tcbk-x4f1},
  url = {https://link.aps.org/doi/10.1103/tcbk-x4f1}
}

@article{Bali:2005fu,
    author = "Bali, Gunnar S. and Neff, Hartmut and Duessel, Thomas and Lippert, Thomas and Schilling, Klaus",
    collaboration = "SESAM",
    title = "{Observation of string breaking in QCD}",
    eprint = "hep-lat/0505012",
    archivePrefix = "arXiv",
    doi = "10.1103/PhysRevD.71.114513",
    journal = "Phys. Rev. D",
    volume = "71",
    pages = "114513",
    year = "2005"
}

@article{Capitani:2018rox,
    author = "Capitani, Stefano and Philipsen, Owe and Reisinger, Christian and Riehl, Carolin and Wagner, Marc",
    title = "{Precision computation of hybrid static potentials in SU(3) lattice gauge theory}",
    eprint = "1811.11046",
    archivePrefix = "arXiv",
    primaryClass = "hep-lat",
    doi = "10.1103/PhysRevD.99.034502",
    journal = "Phys. Rev. D",
    volume = "99",
    number = "3",
    pages = "034502",
    year = "2019"
}

@article{Joswig2023,
  title = {pyerrors: A {P}ython framework for error analysis of {M}onte {C}arlo data},
  volume = {288},
  ISSN = {0010-4655},
  url = {http://dx.doi.org/10.1016/j.cpc.2023.108750},
  DOI = {10.1016/j.cpc.2023.108750},
  journal = {Computer Physics Communications},
  publisher = {Elsevier BV},
  author = {Joswig,  Fabian and Kuberski,  Simon and Kuhlmann,  Justus T. and Neuendorf,  Jan},
  year = {2023},
  month = jul,
  pages = {108750}
}

@article{Wolff2004,
  title = {Monte {C}arlo errors with less errors},
  volume = {156},
  ISSN = {0010-4655},
  url = {http://dx.doi.org/10.1016/S0010-4655(03)00467-3},
  DOI = {10.1016/s0010-4655(03)00467-3},
  number = {2},
  journal = {Computer Physics Communications},
  publisher = {Elsevier BV},
  author = {Wolff,  Ulli},
  year = {2004},
  month = jan,
  pages = {143–153}
}

@article{Wolff2007,
  title = {Erratum to “{M}onte {C}arlo errors with less errors” [{C}omput. {P}hys. {C}omm. 156 (2004) 143–153]},
  volume = {176},
  ISSN = {0010-4655},
  url = {http://dx.doi.org/10.1016/j.cpc.2006.12.001},
  DOI = {10.1016/j.cpc.2006.12.001},
  number = {5},
  journal = {Computer Physics Communications},
  publisher = {Elsevier BV},
  author = {Wolff,  Ulli},
  year = {2007},
  month = mar,
  pages = {383}
}

@article{Ramos2019,
  title = {Automatic differentiation for error analysis of {M}onte {C}arlo data},
  volume = {238},
  ISSN = {0010-4655},
  url = {http://dx.doi.org/10.1016/j.cpc.2018.12.020},
  DOI = {10.1016/j.cpc.2018.12.020},
  journal = {Computer Physics Communications},
  publisher = {Elsevier BV},
  author = {Ramos,  Alberto},
  year = {2019},
  month = may,
  pages = {19–35}
}

@article{Schaefer2011,
  title = {Critical slowing down and error analysis in lattice {QCD} simulations},
  volume = {845},
  ISSN = {0550-3213},
  url = {http://dx.doi.org/10.1016/j.nuclphysb.2010.11.020},
  DOI = {10.1016/j.nuclphysb.2010.11.020},
  number = {1},
  journal = {Nuclear Physics B},
  publisher = {Elsevier BV},
  author = {Schaefer,  Stefan and Sommer,  Rainer and Virotta,  Francesco},
  year = {2011},
  month = apr,
  pages = {93–119}
}

\end{document}